\documentclass[aps,prl,showpacs,preprintnumbers,twocolumn,superscriptaddress]{revtex4-1}
\usepackage{amsmath,amssymb}
\usepackage{bm}
\usepackage{tipa}
\usepackage{upgreek}
\usepackage{comment}
\usepackage{mathrsfs}
\usepackage{graphicx}
\usepackage{braket}
\usepackage{enumitem}
\usepackage{mathbbol}
\usepackage{gensymb}
\usepackage[normalem]{ulem}
\usepackage{color}
\usepackage[colorlinks,bookmarks=true,citecolor=blue,linkcolor=red,urlcolor=blue]{hyperref}
\usepackage{hyperref}

\begin{document}

\title{Spectral response of disorder-free localized lattice gauge theories}

\author{Nilotpal Chakraborty}
\thanks{Corresponding author}
\email{nilotpal@pks.mpg.de}
\affiliation{Max-Planck-Institut f\"{u}r Physik komplexer Systeme, N\"{o}thnitzer Stra\ss e 38, Dresden 01187, Germany}

\author{Markus Heyl}
\affiliation{Max-Planck-Institut f\"{u}r Physik komplexer Systeme, N\"{o}thnitzer Stra\ss e 38, Dresden 01187, Germany}
\affiliation{Theoretical Physics III, Center for Electronic Correlations and Magnetism, Institute of Physics, University of Augsburg, D-86135 Augsburg, Germany}

\author{Petr Karpov}
\affiliation{Max-Planck-Institut f\"{u}r Physik komplexer Systeme, N\"{o}thnitzer Stra\ss e 38, Dresden 01187, Germany}

\author{Roderich Moessner}
\affiliation{Max-Planck-Institut f\"{u}r Physik komplexer Systeme, N\"{o}thnitzer Stra\ss e 38, Dresden 01187, Germany}

\begin{abstract}
We show that certain lattice gauge theories exhibiting disorder-free localization have a characteristic response in spatially averaged spectral functions: a few sharp peaks combined with vanishing response in the zero frequency limit. This reflects the discrete spectra of small clusters of kinetically active regions formed in such gauge theories when they fragment into spatially finite clusters in the localized phase due to the presence of static charges. We obtain the transverse component of the dynamic structure factor, which is probed by neutron scattering experiments, deep in this phase from a combination of analytical estimates and a numerical cluster expansion. We also show that local spectral functions of large finite clusters host discrete peaks whose positions agree with our analytical estimates. Further, information spreading, diagnosed by an unequal time commutator, halts due to real space fragmentation. Our results can be used to distinguish the disorder-free localized phase from conventional paramagnetic counterparts in those frustrated magnets which might realize such an emergent gauge theory.
\end{abstract}

\maketitle     

Lattice gauge theories emerge as low-energy effective theories in a large class of models relevant to quantum many-body physics, especially in frustrated magnets \cite{fradkin2013field,moessner2021topological}. For example, in quantum spin ice and hardcore dimer models, local constraints (such as the ice rule of two spins pointing in and two pointing out locally) yield a 3+1 dimensional compact U(1) lattice gauge theory \cite{Moessner3D,Hermele2004}. The resulting algebraic spin correlations are, strictly speaking, only present at zero temperature, in the absence of  charges (i.e., violations of the constraints). However, perhaps somewhat counterintuitively, one can also find new phases by creating a high density of such charges, provided that these are static. The concomitant sectors of the lattice gauge theories have been shown to exhibit disorder-free localization in one \cite{Smith2017,Brenes2018} and two dimensions \cite{KarpovPRL,Chakrabortyloctrans}. It thus becomes necessary to distinguish disorder-free localization from a paramagnet, the `standard' disordered phase, leading to the twin questions of to what extent one can access novel phases at a finite density of static charges, and above all, what are the distinguishing characteristics of the localized phases?

\begin{figure*}
    \centering
    \includegraphics[width = 17cm, height = 5 cm]{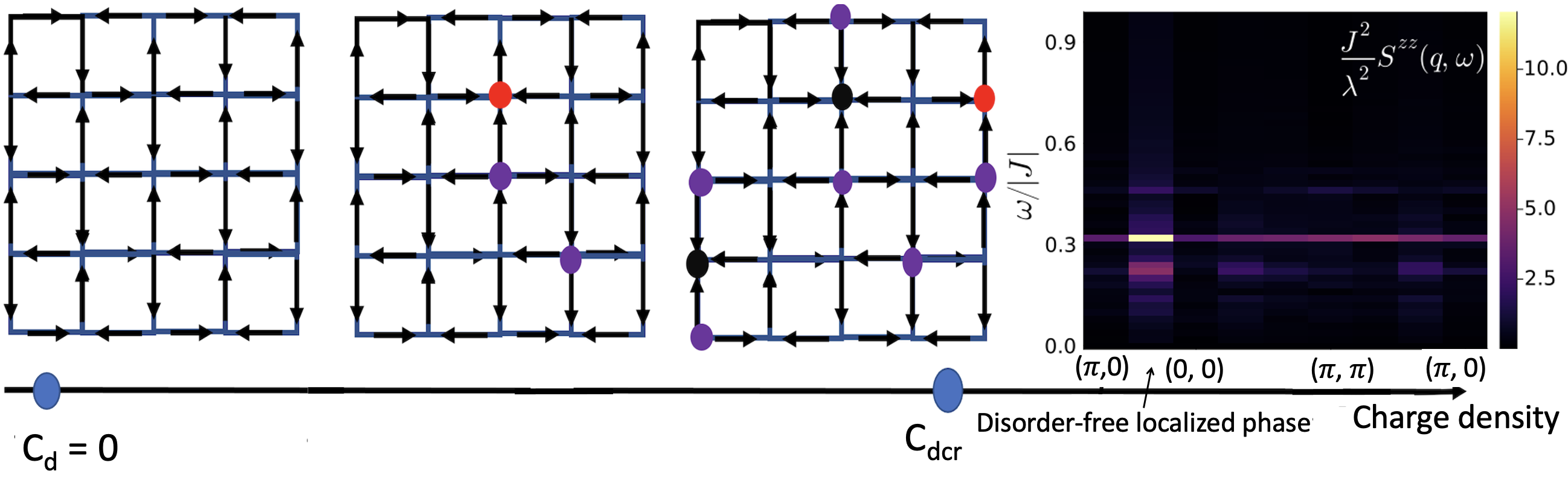}
    \caption{Accessing novel phases by {\it increasing} charge density (colored circles: red, purple and black for $q_r = -1,1$ and $-2$,the arrows $\rightarrow$ and $\downarrow$ represent spin $\ket{\uparrow}$, whereas $\leftarrow$ and $\uparrow$ represent $\ket{\downarrow}$). In the presence of a sufficient density of static charges, one can access the disorder-free localized (DFL) phase ($\rm{C.D} > C_{\rm{dcr}}$). Rightmost panel: Sharp spectral response in spatially averaged spectral functions (in this case, the transverse component of the dynamical structure factor for $|\lambda/J| = 0.2$) with $S^{zz}(q,\omega) \rightarrow 0$ as $\omega \rightarrow 0$, in the disorder-free localized phase.}
    \label{fig1}
\end{figure*}

Here, we present a spectral response dominated by discrete sharp peaks, combined with vanishing weight in the zero frequency limit, as a characteristic signature of the disorder-free localized phase. We show that such a response arises due to formation of local motifs with discrete spectra in such lattice gauge theories, a feature absent in conventional paramagnets which behave diffusively at high temperatures \cite{ConlonChalker}. We numerically obtain the spectral functions by calculating quantum dynamics of the 2+1D U(1) quantum link model (QLM) via a controlled cluster expansion method which allows us to access the dynamics deep in the localized phase. Our conclusions are applicable to a large class of lattice gauge theories with discrete degrees of freedom and static charges, of which the U(1) QLM is a prototypical example.

Besides the aforementioned result we also gain the following insights: i) Disorder-free localized lattice gauge theories provide an example of information freezing due to real-space fragmentation in 2D. ii) In different regimes of the U(1) QLM, there is a difference between local and global response of spatially averaged spectral functions, such a difference could be measured in experiments via local spectroscopy and serves as additional evidence for disorder-free localization physics. There has also been a recent surge in proposals for disorder-free localization \cite{IreneDFL,LiPolar,ZhuHeyl,HodsonDFL}, as well as in proposals for realizing lattice gauge theories in various quantum simulators \cite{schweizer2019floquet,yang2020observation,zhou2022thermalization,semeghini2021probing} and stabilizing disorder-free localization on such platforms \cite{halimeh2021stabilizing}. Hence, distinct experimental signatures of disorder-free localization are of great current relevance.

\textit{Quantum Link model and percolation}: The quantum link model (QLM) is a discrete version of Wilsonian lattice gauge theories also occurring in gauge magnets \cite{chandrasekharan1997quantum,horn1981finite,orland1990lattice,Wiese2013}. Several models in condensed matter physics, for example the toric code \cite{kitaev2003fault}, quantum dimer and ice models \cite{RokhsarPRL,Moessnerising,Shannoncyclic} can be cast as  quantum link models. We use the 2+1 D U(1) QLM with
\begin{equation}
    \hat{H} = J \sum_{\square} (\hat{U}_{\square} + \hat{U}_{\square}^{\dagger}) - \lambda \sum_{\square} (\hat{U}_{\square} + \hat{U}_{\square}^{\dagger})^2  ,
    \label{eq:HamiltonianQLM}
\end{equation}
where $\hat{U}_{\square} = \hat{S}^{+}_{\textbf{r},i} \hat{S}^{+}_{\textbf{r}+\textbf{i},j} \hat{S}^{-}_{\textbf{r}+\textbf{j},i} \hat{S}^{-}_{\textbf{r},j}$ flips all spins on a plaquette if they satisfy a certain orientation (clockwise/anticlockwise in Fig.~\ref{fig1}). Here, $\hat{S}^{+/-}_{\textbf{r},\mu}$ denotes the spin raising/lowering operators for the spin joining sites $\textbf{r}$ and $\textbf{r} + \hat{\mu}$, where $\hat{\mu} = \hat{i}/\hat{j}$ is a lattice vector. The first term in Eq. (\ref{eq:HamiltonianQLM}) is a kinetic energy term, whereas the second is a potential energy term \cite{RokhsarPRL}. Spin configurations are characterized by the set of static charges on all sites of the lattice and the charge on each site $q_{\textbf{r}}$ is determined by the eigenvalue equation of the generator of gauge transformations (which evaluates the difference in incoming and outgoing arrows from a site, see Fig. \ref{fig1}), $G_r \ket{\psi} = q_r \ket{\psi}$ \cite{Brenes2018}. For a spin-$1/2$ Hamiltonian $q_{\textbf{r}} \in  \{-2,-1,0,1,2\}$.
This model has been shown to host disorder-free localized physics and a continuous localization-delocalization transition \cite{KarpovPRL,Chakrabortyloctrans}. Crucially, the charges are static in our model. \\
 
We consider the problem with Eq. (\ref{eq:HamiltonianQLM}) as the effective Hamiltonian. Without any static background charges we get square ice, i.e. the local constraint of two in and two out spins on every vertex.  On increasing charge density beyond a critical density $C_{\rm{dcr}}$, there is a disorder-free localized phase \cite{KarpovPRL,Chakrabortyloctrans}. If one thinks of these charges as being thermally activated, the disorder-free localized phase would correspond to a high temperature phase (recently proposed in a different context for 1D \cite{halimeh2022temperature}). Temperature-dependent dynamics for gauge theories has also been studied in context of the Kitaev model \cite{Brenigtemperature}.\\
Our disorder-free localization can be studied using a percolation model, where one starts from a classical random spin configuration, which fixes the initial charge configuration, and then tracks all possible spin configurations induced by single plaquette flips (1st term in Eq. (\ref{eq:HamiltonianQLM})). Plaquette dynamics in such a model is constrained by the static charge environment and certain charge configurations will render a subset of plaquettes permanently frozen. If the initial state hosts a sufficient density of background charges to fragment the lattice into disconnected finite clusters of kinetically active plaquettes, localization ensues (see \cite{KarpovPRL,Chakrabortyloctrans} for percolation procedure).

\textit{Effective exact diagonalization using cluster expansion in the localized phase:} We employ an effective exact diagonalization procedure to calculate the quantum dynamics of our 2D interacting system deep in the localized phase. The percolation model gives us the real-space cluster structure, which in the localized phase comprises only finite disconnected clusters. In contrast, in the delocalized phase there is an infinite percolating cluster coexisting with the finite clusters.\\ The decomposition of the whole lattice into only finite clusters in the localized phase makes the quantum problem numerically tractable \cite{Chakrabortyloctrans}.
To calculate spatially averaged expectation values of time-dependent correlation functions for the lattice, we thus use a cluster expansion approach. We calculate the quantum dynamics individually for each cluster using exact diagonalization. We then sum the results for all clusters upto a certain size after which we find that including larger clusters adds negligible weight to our observables. Plaquettes in a cluster are connected via subsequent single plaquette flips, whereas plaquettes belonging to different clusters (disjoint in real space) are uncorrelated. Since we calculate dynamical expectation values of two point correlators, the only non-zero contributions come from pairs of spins belonging to the same cluster. Note that our cutoff procedure works only deep in the localized phase where the mean cluster size of the corresponding non-percolating phase is not too large. Closer to the transition, the mean cluster size increases and the contribution from finite but large clusters becomes important, hence the problem goes beyond the realm of our exact diagonalization numerics. \\
Frozen spins also contribute a delta function response exactly at zero frequency. However, since our results pertain to non-zero frequencies, frozen spins are only included in the denominator in Eqs. (\ref{uneqcomm})(1st equation) and (\ref{spec}) for calculating global spectral functions.\\
Besides the global response, we also calculate local spectral functions, spatially averaged over a finite size cluster. To do so we average over all spins belonging to all clusters of a particular size, over multiple disorder realizations.

\begin{figure}
    \centering
    \includegraphics[width = 8cm, height = 14cm]{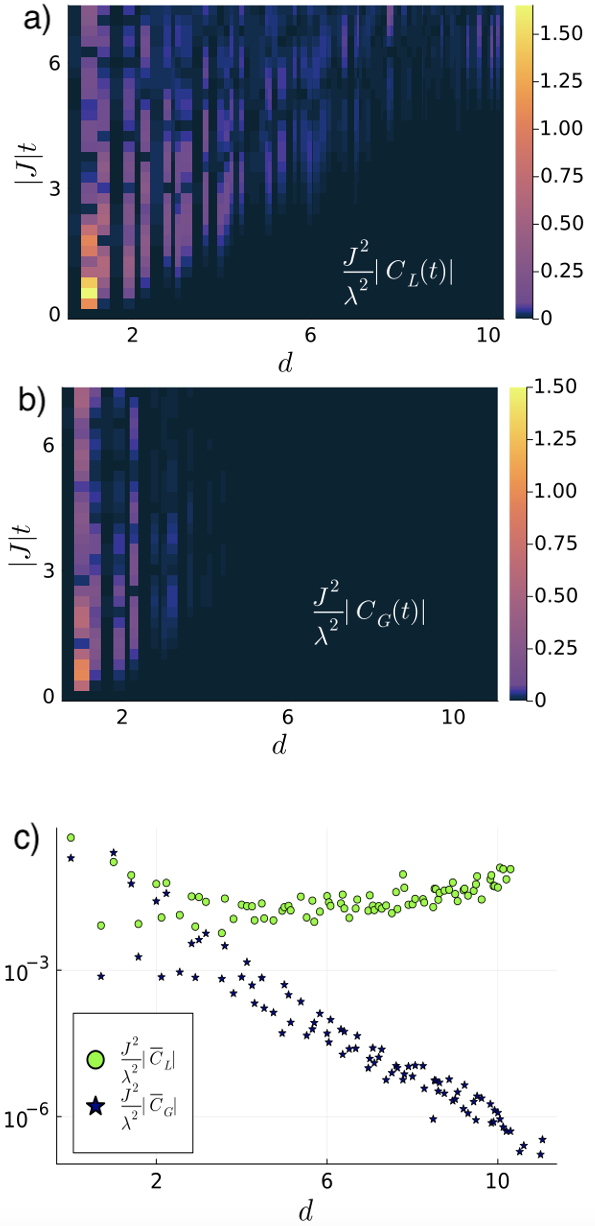}
    \caption{Local and global information spreading quantified by $J^2/\lambda^2 |C_{L/G}(t)|$ for $|\lambda/J| = 0.2$. a) Information spreading within a 18 plaquette cluster b) Information freezing in the full system c) Long time average vs distance for a) and b).}
    \label{fig2}
\end{figure}

\textit{Information freezing due to fragmentation}:
We use our method to first calculate unequal time commutators which are related to the susceptibility via the Kubo formula. Such commutators (also their relatives, the out-of-time-ordered correlators) quantify information spreading, a quantity related to entanglement growth, and have been a measure of chaos and Hilbert-space fragmentation in quantum many-body systems \cite{maldacena2016bound,Salafrag,Khemanishatter,Keyserlinkotoc}. We calculate
\begin{equation} \label{uneqcomm}
\begin{split}
    &C_G(d,t) = \frac{1}{N}\sum_{\substack{\textbf{r}_i,\textbf{r}_j \in \mathcal{L} \\ |\textbf{r}_i-\textbf{r}_j| = d}} \bra{\psi} \big[\hat{S^z_{\textbf{r}_i}}(t),\hat{S^z}_{\textbf{r}_j}(0)\big] \ket{\psi} \\
        &= \frac{1}{N_{d}}\sum_{|\rm{cl}| <= P_c} \sum_{\substack{\textbf{r'}_i,\textbf{r'}_j \in \rm{cl} \\ |\textbf{r'}_i-\textbf{r'}_j| = d}} \bra{\psi_{\rm{cl}}} [\hat{S^z_{\textbf{r'}_i}}(t),\hat{S^z}_{\textbf{r'}_j}(0)] \ket{\psi_{\rm{cl}}}\\
     &C_L(d,t) =   \frac{1}{N'_{d}}\sum_{|\rm{cl}| = P} \sum_{\substack{\textbf{r'}_i,\textbf{r'}_j \in \rm{cl} \\ |\textbf{r'}_i-\textbf{r'}_j| = d}} \bra{\psi_{\rm{cl}}} [\hat{S^z_{\textbf{r'}_i}}(t),\hat{S^z}_{\textbf{r'}_j}(0)] \ket{\psi_{\rm{cl}}}
\end{split}
\end{equation}
where $\hat{S^z}_{\textbf{r}_i}$ and $\hat{S^z}_{\textbf{r}_j}$ are spins at $\textbf{r}_i$ and $\textbf{r}_j$, $|\textbf{r}_i - \textbf{r}_j| = d$, $\ket{\psi_{cl}}$ is a Haar random state drawn from the Hilbert space spanned by all possible configuration states of the cluster $\rm{cl}$ and $N_{d}$ is the number of pairs of spins in the lattice $\mathcal{L}$ separated by distance $d$, excluding the pairs of spins that belong to all clusters of size $|cl| > P_c$ as per our method ($P_c = 18$ for this paper). $N'_{d}$ is the number of pairs of spins separated by distance $d$ in all clusters with $P$ plaquettes. \\ 
We analyze global information spreading represented by $C_G(d,t)$ in Eq. (\ref{uneqcomm}), related to entanglement growth in the full system as well as local information spreading $C_L(d,t)$, related to the same in a finite size cluster.\\
Fig. \ref{fig2}a and b display a stark difference between the two cases in the $|\lambda/J| \ll 1$ regime. There is a signal up to large distances with a light cone like structure for information spread within a cluster, affirming the ergodic nature of large finite size clusters in this regime \cite{Chakrabortyloctrans}. However, for the whole lattice, there is an exponential decay with a correlation length of around 3 to 4 lattice spacings, i.e. next-nearest-neighbour plaquettes, implying information freezing and saturation of entanglement growth. Such a qualitative difference is even more apparent in the long-time average of the commutators with distance in Fig. \ref{fig2}c. Note that such information freezing is present for all values of $|\lambda/J|$ and is a generic feature of all disorder-free localized lattice gauge theories, where kinetic constraints induced by static charges, cause real-space fragmentation.

\begin{figure}[t]
    \centering
    \includegraphics[scale = 0.52]{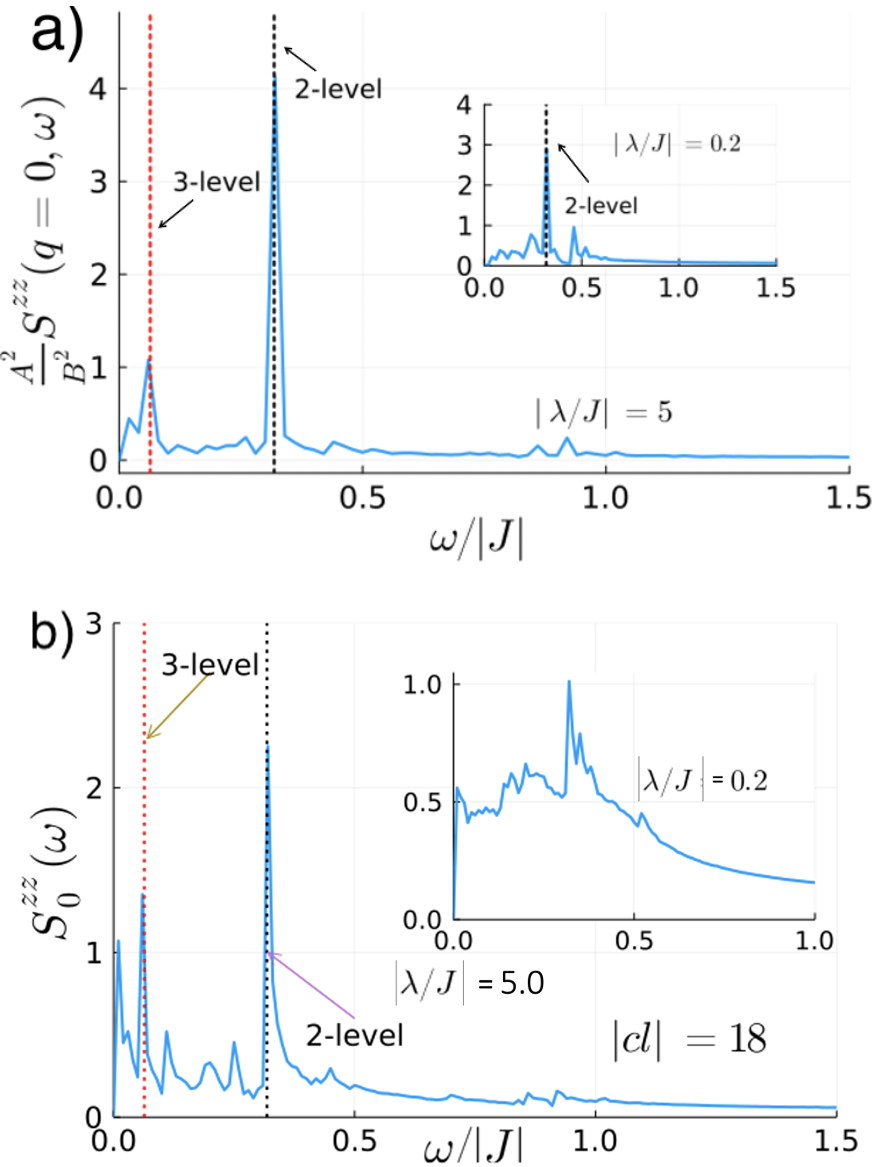}
    \caption{ Evidence of disorder-free localization from high temperature ($T = 5 \text{max}(|\lambda|,|J|)$) spatially averaged spectral functions. a) The $q=0$ component of the transverse component of the dynamic structure factor, A = max($|\lambda|$,$|J|$) and B = min($|\lambda|$,$|J|$), cutoff size $P_c = 18$ and $|\lambda/J| = 5.0$ , (inset) same quantity as main plot, but for $|\lambda/J| = 0.2$  b) On-site spectral function averaged over all spins belonging to all 18 plaquette clusters for $|\lambda/J| = 5.0$ and (inset) $|\lambda/J| = 0.2$}
    \label{fig3}
\end{figure}

\textit{High temperature spectral functions}
We also calculate high-temperature ($T \gg \rm{max}(|\lambda|,|J|)$) spectral functions, which are relevant for neutron scattering experiments on frustrated magnets that might realize a U(1) QLM type effective theory. To do so, one employs arguments from dynamical quantum typicality (DQT) \cite{Robintypical,Elsayedtypical} which state that a pure state drawn from a Haar-random distribution gives the same expectation values as the entire statistical ensemble. One can then define correlation functions as
\begin{multline}
\begin{split}
    S^{zz}&(d,t,T)= \frac{1}{N}\sum_{\substack{\textbf{r}_i,\textbf{r}_j \in \mathcal{L} \\ |\textbf{r}_i-\textbf{r}_j| = d} } \dfrac{ \bra{\Phi_{\beta(t)}} \hat{S^{z}_{\textbf{r}_i}} \ket{\varphi_{\beta(t)}}}{\braket{\Phi_{\beta}(0)|\Phi_{\beta}(0)}}
    + \epsilon \\
    &=\frac{1}{N'_{d}} \sum_{\substack{|cl| \\ <= P_c}} \sum_{\substack{\textbf{r}_i,\textbf{r}_j \in cl \\ |\textbf{r}_i-\textbf{r}_j| = d} } \dfrac{ \bra{\Phi_{\beta_{cl}}(t)} \hat{S^{z}_{\textbf{r}_i}} \ket{\varphi_{\beta_{cl}}(t)}}{\braket{\Phi_{\beta_{cl}}(0)|\Phi_{\beta_{cl}}(0)}} 
    + \epsilon
\end{split}
    \label{spec}
\end{multline}
where $\ket{\Phi_{\beta_{cl}}(t)} = e^{-iH_{cl} t - \beta H_{cl}/2} \ket{\psi_{cl}}$, $\ket{\varphi_{\beta_{p}}(t)} = e^{-iH_{cl} t}\hat{S^{z}_{\textbf{r}_j}} e^{- \beta H_{cl}/2} \ket{\psi_{cl}}$,$\beta = 1/T$, $H_{cl}$ is the Hamiltonian in Eq. (\ref{eq:HamiltonianQLM}) but with the sum over all plaquettes belonging to cluster $cl$, and
$\epsilon$ denotes the error which is exponentially small in the number of thermally occupied eigenstates. The typicality approach also works very well in a localized phase for large Hilbert spaces \cite{Brenigtypical}.\\
Taking the space-time Fourier transform of Eq. (\ref{spec}), gives the transverse component of the dynamical structure factor. As shown in Figs. \ref{fig1} and \ref{fig3}a, there are dominant delta-function like peaks at low frequencies amidst a background of featureless response and a vanishing response as $\omega \rightarrow 0$. This is the hallmark of disorder-free localization and distinguishes it from paramagnetic response \cite{ConlonChalker}. Such response remains robust on increasing system size and for all values of $\lambda$ and $J$ since the dominant peaks arise from the formation of small clusters (see next section) deep in the localized phase due to fragmentation induced by the static charges. The density of such clusters saturates at a constant value as a function of background charge density deep in the localized phase, smoothly decays across the localization transition and then vanishes as a polynomial deep in the delocalized phase, as in standard percolation. Crucially, on crossing the localization transition, an infinite ergodic cluster emerges which has a continuous spectral response \cite{Chakrabortyloctrans}. Hence, on crossing the transition point from the localized side, there is a shift in weight from the discrete peaks to the continuous response.\\ 
However, the infinite cluster exhibits long-time hydrodynamic behaviour with U(1) charge conservation. Assuming diffusion in 2D, i.e. a temporal decay of $t^{-d/2} = 1/t$, one would expect a logarithmic divergence of the spectral response as $\omega \rightarrow 0$. The localized phase, however, has no such contribution. Moreover, since the localization physics is dictated by the small clusters, there is a vanishing spectral response as $\omega \rightarrow 0$ due to their discrete and non-degenerate spectra.

Besides the above global signatures, local spectral functions, such as the temporal Fourier transform of onsite correlators, $S^{zz}_0 (\omega)$, spatially averaged over spins belonging to all clusters of a particular size, evolves from a smooth curve with short bumps in the $|\lambda/J| \ll 1$ regime to a discrete set of peaks in the $|\lambda/J| \gg 1$ regime, as in Fig. \ref{fig3}b.  Such a difference indicates a crossover from ergodic to non-ergodic behaviour within large but finite clusters as predicted in Ref.~\cite{Chakrabortyloctrans}.

\textit{Analytic estimate of dominant peaks in spectral functions:}
The position of the discrete peaks observed in the spectral functions above can be accurately estimated from the local physics of clusters of few plaquettes which form two and three level systems. Consider a two level system, for example a single flippable plaquette, the eigenvalues of such a system will be $-\lambda + J$ and $-\lambda - J$. Such a system should have a discrete peak at $\omega = J/ 
\pi$. Similarly, for a two-plaquette system with 3 states ($11/10/01$; $1$ for flippable and $0$ for not flippable), the transition between the closest spaced pair is given by a discrete peak at $\omega = J^2/(\lambda \pi)$ in the $|\lambda/J| \gg 1$ regime. The dominance of these peaks arising from such local physics implies: i) On the global scale the dominant response deep in the localized phase is from the small clusters and ii) In the $|\lambda/J| \gg 1$ regime, the physics of larger clusters is effectively dominated by local motifs which form two- or three-level systems. Since within a cluster there is no real-space fragmentation, the local motifs arise purely due to interference effects. However, this apparent non-ergodicity is a finite-size effect with dominant discrete peaks over a large finite-size regime, the spectral weight of which eventually decreases on increasing size giving way to a continuous spectrum in the infinite-size limit~\cite{Chakrabortyloctrans}. The former contribution, however, remains discrete in the infinite-size limit, as the presence of a finite density of charges will always induce real-space fragmentation and form small clusters. These clusters have the same spectra irrespective of their position, hence even after spatial averaging there remains a sharp response in all regimes of the Hamiltonian.\\
\textit{Discussion and outlook:} In this work we have presented numerical results for spatially averaged spectral functions in 2D as characteristic evidence for disorder-free localization. Local spectroscopic signatures have also been proposed for conventional disordered MBL, as a way to distinguish between strong and weak MBL \cite{nandkishorespec,johrispecmbl}. However, most sharp spectral features are blurred out on spatial averaging, with a soft spectral gap at zero frequency being the only distinct consequence of both kinds of MBL in the presence of random disorder. For the disorder-free localized phase, there are dominant sharp peaks in the dynamical structure factor as well as vanishing response as $\omega \rightarrow 0$, both of these can, in principle, be observed in spectroscopic experiments. Our work can be translated to higher dimensions in which such sharp spectral features could serve as evidence for quantum spin ice materials in 3D, where the analog of the charges would be static defects. Our results on the unequal-time commutators further motivate lattice gauge theories, possibly realized in quantum simulators, as a fertile platform to probe information spreading and chaos. The problem of lattice gauge theory with mobile matter adds a dynamic charge component to the problem, hence an important future direction would be to explore how, in that case, the signatures presented in this work would be modified.

\textit{Acknowledgements.---}
We thank Hongzheng Zhao and Adam McRoberts for valuable discussions. This project has received funding from the European Research Council (ERC) under the European Union’s Horizon 2020 research and innovation programme (Grant Agreement No. 853443), and M.H. further acknowledges support by the Deutsche Forschungsgemeinschaft via the Gottfried Wilhelm Leibniz Prize program. This work was in part supported by the Deutsche Forschungsgemeinschaft under grants SFB 1143 (project-id 247310070) and the cluster of excellence ct.qmat (EXC 2147, project-id 390858490). 

\bibliography{Bib2.bib}
\end{document}